# Pressure-induced Superconductivity in Sulfur-doped SnSe Single Crystal Using Boron-doped Diamond Electrode-prefabricated Diamond Anvil Cell

Ryo Matsumoto[a,b], Hiroshi Hara[a,b], Hiromi Tanaka[c], Kazuki Nakamura[c], Noriyuki Kataoka[c,d], Sayaka Yamamoto[c], Aichi Yamashita[a,b], Shintaro Adachi[a], Tetsuo Irifune[e], Hiroyuki Takeya[a], and Yoshihiko Takano[a,b]

[a]National Institute for Materials Science, 1-2-1 Sengen, Tsukuba, Ibaraki 305-0047, Japan
[b]University of Tsukuba, 1-1-1 Tennodai, Tsukuba, Ibaraki 305-8577, Japan
[c]National Institute of Technology, Yonago College, 4448 Hikona, Yonago, Tottori 683-8502, Japan
[d]Okayama University, 3-1-1, tsushimanaka, kitaku, Okayama 700-8530, Japan
[e]Geodynamics Research Center, Ehime University, Matsuyama, Ehime 790-8577, Japan

**Abstract**

Sulfur-doped SnSe single crystal was successfully synthesized using a melt and slow-cooling method. The chemical composition and valence state of the obtained sample were analyzed by X-ray photoelectron spectroscopy. The pressure range of a diamond anvil cell with boron-doped diamond electrodes was upgraded to 104 GPa using nano-polycrystalline diamond anvil to investigate a pressure effect for the sample. Electrical resistivity measurements of sulfur-doped SnSe single crystal showed the insulator-metal-superconductor transition by applying high pressure up to 75.9 GPa.

## 1. Introduction

Recently, a lot of high performance functional materials are designed from a view point of an electric band structure. A specific band structures of "flat band" near a Fermi level, such as multivalley [1], pudding-mold [2], and topological-type [3] structures are especially spotlighted in the development of thermoelectric and superconducting materials. If such kinds of flat band approach Fermi level, thermoelectric properties of electrical conductivity and Seebeck coefficient would be enhanced [2,4]. If the flat band crosses the Fermi level, superconductivity would be realized due to high density of states (DOS) [5-7].

It is suggested by angle-resolved photoemission spectroscopy that SnSe has such a flat band of multi-valley type near the Fermi level [8]. This compound shows remarkable thermoelectric property of ultrahigh figure of merit ZT value of 2.6 at 923 K, due to its low thermal conductivity, resistivity, and high Seebeck coefficient [9,10]. If the flat band of SnSe crosses the Fermi level, it was also expected to appear the superconductivity with high transition temperature ($T_c$).

It is known that the SnSe exhibits structural phase-transition from stable phase at ambient pressure to NaCl-type and CsCl-type structures under high pressure [11-12]. Even in these structures, SnSe has the flat band near the Fermi level [12-14]. The CsCl-type SnSe shows coexistence of nontrivial topology and superconductivity under high pressure [12]. According to some theoretical studies [15,16], Dirac line nodes suggest existence of topologically nontrivial surface state, called

"flat band", which is considered as a way to induce high-$T_c$ superconductivity. SnSe is ideal compound to confirm the strategy for inducing high-$T_c$ superconductivity using the specific band structure "flat band".

Hence, the functionality of SnSe would be dramatically changed by tuning the position of Fermi level because the all crystal structures in this compound provide high DOS near the Fermi level, which are suitable for thermoelectric and superconducting properties. The methods to tune the position of Fermi level are carrier doping and/or high pressure application. For example, an existence of high DOS near the Fermi level was suggested by a band calculation in highest $T_c$ superconductor $H_3S$ [6]. Theoretical studies indicate this superconductivity is enhanced up to 280 K by carrier doping using elemental substitution and applying much higher pressure of 250 GPa [7]. Considering the similarity of band structure in $H_3S$, SnSe is also a candidate high-$T_c$ superconductor by tuning the position of the Fermi level.

In this study, we focused on chemical and physical pressure effects for SnSe to adjust the band structure. If S substitutes the Se site, the lattice constant will be shrunk. It is namely the chemical pressure application against the crystal. Moreover, we directly applied the physical pressure to the sulfur-doped SnSe by using a boron-doped diamond electrode-prefabricated diamond anvil cell (BDDE-DAC) high pressure apparatus [17-19]. In-situ electrical resistivity measurements were performed under high pressure up to 75.9 GPa. The band edge with high DOS in SnSe can approach to the Fermi level by applying the pressure due to an increase of bandwidth.

## 2. Experimental

Single crystals of sulfur-doped SnSe were grown by a melt and slow-cooling method. Starting materials of Sn grains (99.99%), Se grains (99.999%) and S grains (99.9999%) were put into an evacuated quartz tube in the stoichiometric composition of $SnSe_{0.99}S_{0.01}$. The ampoule was heated at 600ºC for 8 hours, subsequently at 930ºC for 10 hours. The obtained powders were ground and loaded into an evacuated quartz tube. The sample was heated again at 930ºC for 10 hours, and slowly cooled to 600ºC for 16 hours followed by furnace cooling.

Chemical composition and valence state of obtained sample was estimated by X-ray photoelectron spectroscopy (XPS) analysis using AXIS-ULTRA DLD (Shimadzu/Kratos) with AlK$\alpha$ X-ray radiation ($hv$ = 1486.6 eV), operating under a pressure of the order of $10^{-9}$ Torr. The analyzed area was approximately 1×1 $mm^2$. The background signals were subtracted by active Shirley method using COMPRO software [20].

It is necessary to develop a measurement technique of electrical transport properties under high pressure in the range of mega-bar region for investigation of pressure effect in the all crystal structures of sulfur-doped SnSe. In this study, a pressure-range of the BDDE-DAC [19] was upgraded by using nano-polycrystalline diamond (NPD) anvil, which has great Vickers hardness [21].

The fabrications processes of boron-doped diamond (BDD) electrodes and undoped diamond (UDD) insulating layer in BDDE-DAC were adjusted for NPD as following procedures. The BDD film was deposited on the NPD anvil by microwave-assisted chemical vapor deposition using $CH_4$ diluted with $H_2$. The total pressure, total gas flow rate, and microwave power during the growth were maintained at 70 Torr, 300 sccm, and 1100 W, respectively. The boron/carbon ratio was adjusted by 2500 ppm using trimethylboron source. After the deposition, a metal mask of Ti/Au was fabricated by using electron beam lithography in combination with a scanning electron microscope (JEOL: JSM-5310) equipped with a nanofabrication system (Tokyo Technology: Beam Draw) on the BDD. The unmasked region was milled by $O_2$ plasma using a reactive ion-etching machine (Sumitomo Precision Products: MUC-21 ASE-SRE). The UDD insulating layer with 200 nm thickness was also fabricated on the NPD anvil with BDD electrodes through a selective growth method by referring the previous reports [19].

The in-situ resistivity measurements of sulfer-doped SnSe single crystal were demonstrated under high pressure using the developed DAC. The cubic boron nitride powders with ruby manometer were used as a pressure-transmitting medium. The applied pressure values were estimated by the fluorescence from ruby powders [22] and the Raman spectrum from the culet of top diamond anvil [23] by an inVia Raman Microscope (RENISHAW).

## 3. Results and discussion

Figure 1(a) shows XPS spectrum of wide scan up to the binding energy of 1400 eV. The detected elements with an orbital of largest ionization cross section were labeled in the spectrum. Sn $3d$ and Se $3d$ orbital peaks were came from the sample crystal, C $1s$ and O $1s$ were from a sample holder. It is difficult to detect the signals of small amount sulfur from photoelectron peaks because the positions of S $2p$ and S $2s$ orbital peaks were situated in quite similar energy region to that of Se $3p$ and Se $3s$ orbital peaks, respectively [24]. To detect sulfur signal, Auger peak of S LMM was measured as shown in the Fig. 1(b). This spectrum clearly showed an existence of small amount sulfur elements in the sample crystal. Much detailed chemical state analysis is required as a further study using S $1s$ orbital peak on higher energy region.

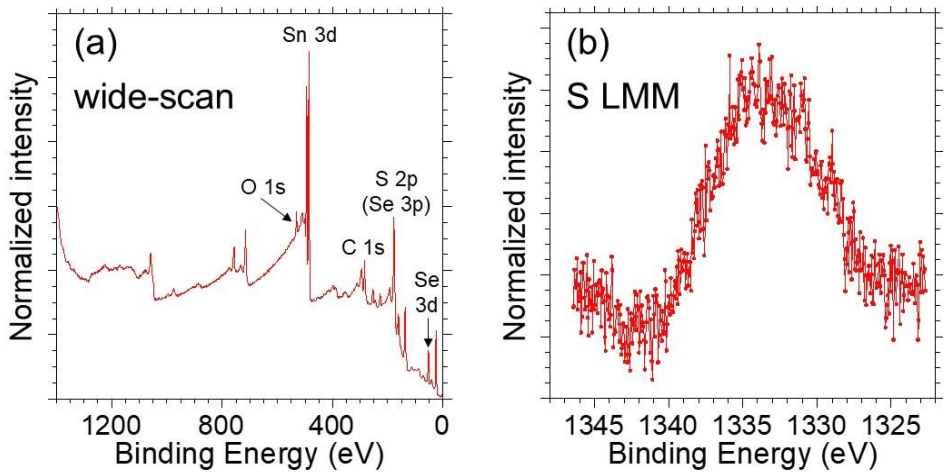

**Figure 1. XPS spectra of obtained sulfur doped SnSe single crystal. (a) spectrum of wide scan up to the binding energy of 1400 eV. (b) S LMM Auger spectrum.**

High-resolution XPS measurements of Sn $3d$ orbital were carried out to investigate a valence state of the obtained sulfur-doped SnSe single crystal. Figure 2(a) showed Sn $3d$ spectrum of cleaved sample using scotch tape in a highly vacuumed pre-chamber in the order of $10^{-7}$ Torr to obtain an intrinsic valence state. In both Sn $3d_{5/2}$ and Sn $3d_{3/2}$ orbitals, sharp peaks were clearly observed at 485.7 eV and 494.1 eV. On the other hand, the spectrum of air-exposed sample for 3 hours after the cleavage showed the other peaks on the left shoulders of the original peaks as shown in Fig. 2(b). The chemical shift of 0.7-1.5 eV between $Sn^{4+}$ and $Sn^{2+}$ was accordingly reported by previous studies [25,26]. On this basis, Sn *3d* peaks could be de-convoluted into two peaks (numbered 1, 2, 3 and 4). Peak 1 (centered at 485.7 eV) and peak 3 (centered at 494.1 eV) correspond to $Sn^{2+}$. The spin–orbit splitting of 8.4 eV is consistent with the data reported in the literature [27]. Peak 2 (centered at 486.9 eV) and peak 4 (centered at 495.3 eV) are ascribed to $Sn^{4+}$. These results indicate that the surface of sulfur-doped SnSe single crystal is quite sensitive for oxygen in the air. According to a previous report [10], undoped SnSe itself is also sensitive against oxygen contamination. Here, the thickness of the $Sn^{4+}$ layer was estimated using the equation, $d=L\cos\theta\ln(I_A/I_B+1)$ [20], where $d$ is the thickness, $L$ is inelastic mean free path of photoelectron from the sample, $\theta$ is emission angle, $I_A$ and $I_B$ are the peak area intensities from the $Sn^{4+}$ and $Sn^{2+}$ peaks, respectively. The value of $L$ is 2.4 nm [28], $\theta$ is 0°, and $I_A/I_B$ is 0.5. The calculation indicates that the thickness of 9.7 Å from the sample surface is the valence state of $Sn^{4+}$, and that of the bulk is $Sn^{2+}$.

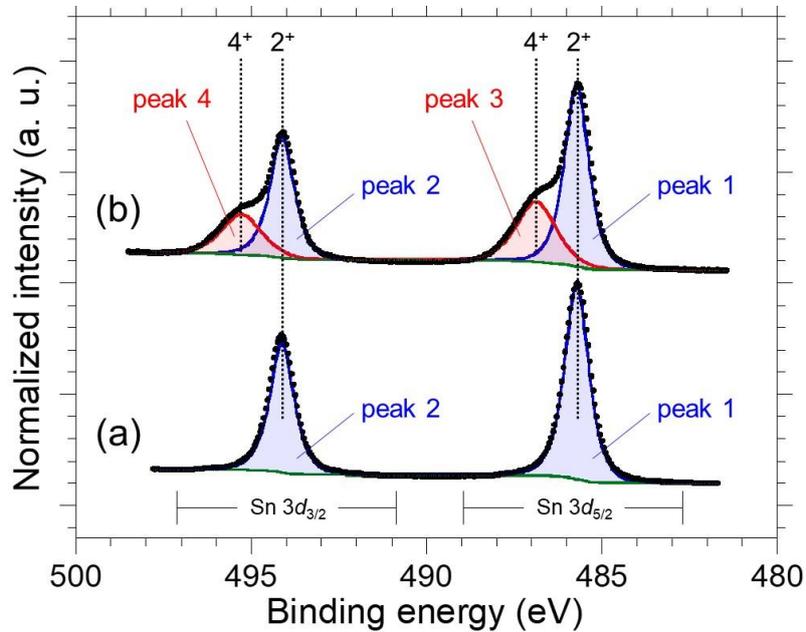

**Figure 2. High-resolution XPS measurements of Sn 3d orbital of (a) cleaved sample in a highly vacuumed pre-chamber in the order of $10^{-8}$ Torr using scotch tape, and (b) air-exposed sample for 3 hours after the cleavage**

The BDDE-DAC with NPD anvil as shown in the schematic image of Fig. 3(a) was assembled for high pressure generation. Here, the temperature dependence of resistance in the BDD on NPD anvil is presented in Fig. 3(b). The resistance rapidly dropped to zero at 2.3 K, corresponding to the superconducting transition. It suggests that the boron concentration in the diamond is ~$10^{21}$ cm$^{-3}$ order [29]. The cleaved sample was put on the BDD electrodes of the developed DAC. The pre-pressed metal gasket of stainless steel sheet with a thickness of 200 μm was fixed on the bottom anvil.

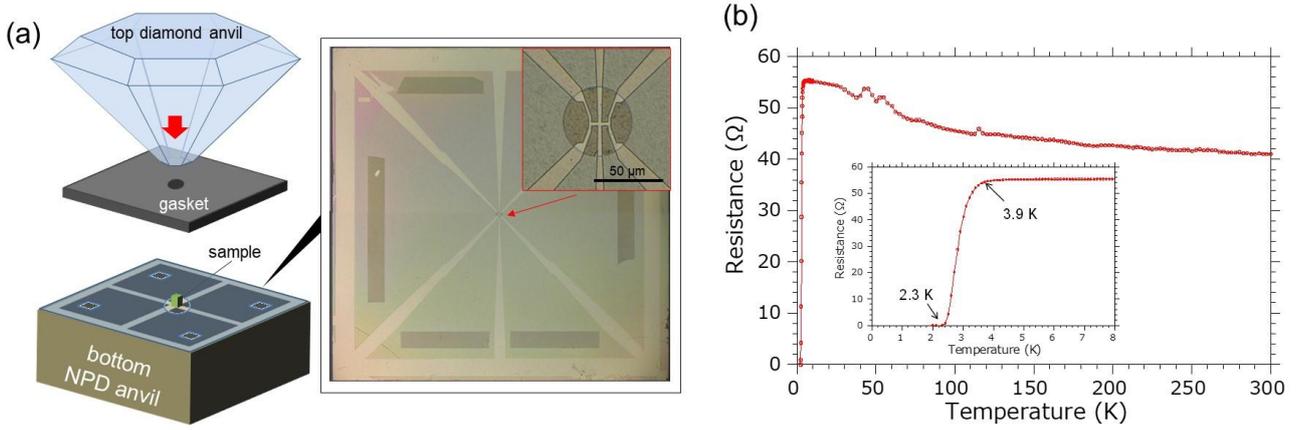

**Figure 3. (a) optical microscope image of the obtained NPD anvil with BDD electrodes covered by UDD insulating layer and schematic image of DAC with flat NPD anvil, (b) temperature dependence of resistance in BDD electrodes on NPD anvil.**

Figure 4 shows the estimated pressures as a function of the stroke length of anvil using BDDE-DAC assembly of (a) flat single crystal diamond anvil with φ0.4 mm culet anvil as a reference. The pressure value tended to saturate from ~50 GPa, and both top and bottom anvils were broken at 65 GPa. Fig.4(b) shows the estimated pressure using flat NPD anvil with φ0.3 mm culet anvil. The pressure value was linearly increased as a function of the stroke length, and the maximum pressure was achieved to 104 GPa.

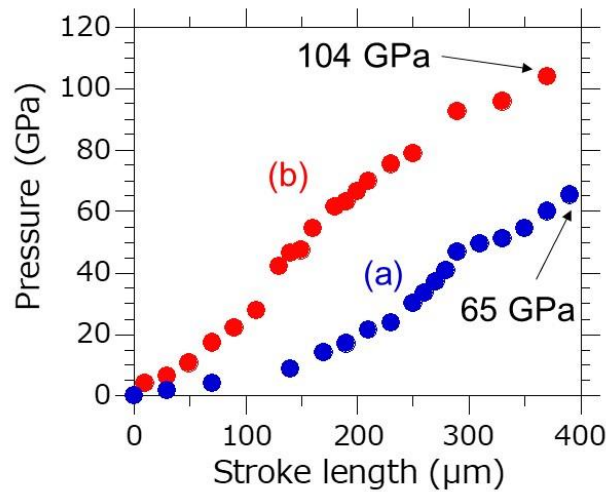

**Figure 4. Estimated pressures as a function of the stroke length of anvil using DAC assembly of (a) flat single crystal diamond anvil with φ0.4 mm culet anvil and (b) flat NPD anvil with φ0.3 mm culet anvil.**

Figure 5(a) shows temperature dependence of resistance in sulfur-doped SnSe single crystal under various pressures. The sample showed semiconducting behavior with band gap of ~200 meV below 10.7 GPa. To increase the pressure above 17.3 GPa, metal-insulator transition was observed. When the pressure increased up to 42.4 GPa, a sharp drop of resistance corresponding to superconductivity was clearly observed at 3.9 K as shown in Fig.5 (b). With increase of pressure, superconducting transition temperature increases, and shows the maximum value of $T_c^{onset}$ ~4.3 K, $T_c^{zero}$ ~3.4 K at 57.5 GPa. Although the $T_c$ slightly tended to decrease after 57.5 GPa, it was kept up to 75.9 GPa. The superconducting transition width became sharper with increase of the pressure.

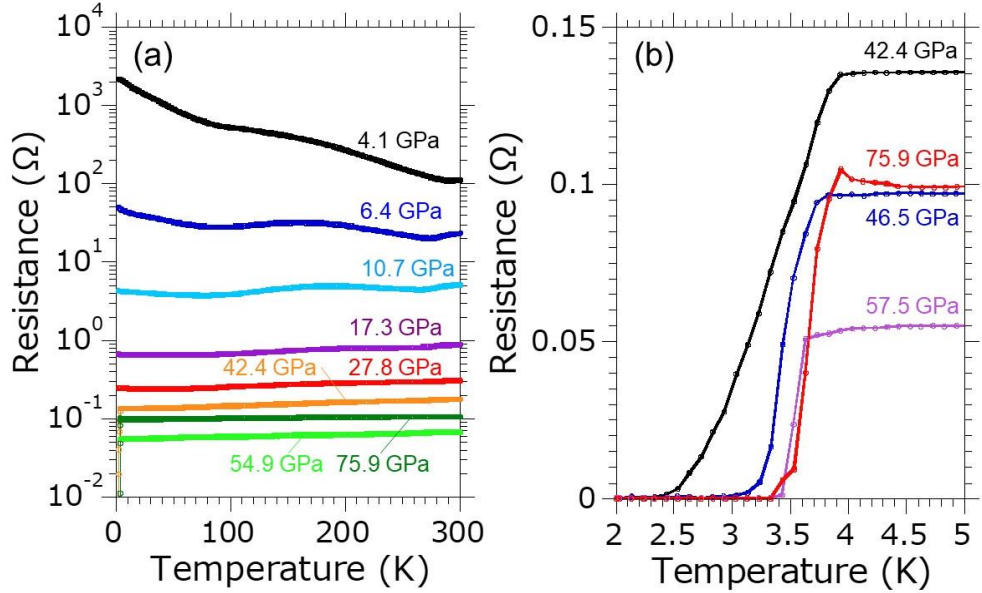

**Figure 5. (a) temperature dependence of resistance in SnSe$_{0.99}$S$_{0.01}$ under various pressures. (b) enlargement of superconducting transition.**

Figure 6 shows a pressure phase diagram of sulfur-doped SnSe single crystal up to 75.9 GPa. The resistance at 300 K and 2 K, $T_c$ at zero resistance ($T_c^{zero}$) and onset $T_c$ ($T_c^{onset}$) were plotted as a function of applied pressures. The sample showed metal-insulator transition at 17.3 GPa, and superconducting transition at 42.4 GPa. The superconductivity still survived at least 75.9 GPa. This pressure-phase diagram basically corresponds to that of the previous report of undoped SnSe [12]. However, higher $T_c^{onset}$ of ~4.3 K and $T_c^{zero}$ of ~3.4 K were observed in our measurement, compared to that of pure SnSe. Moreover, although the $T_c$ of undoped SnSe was rapidly decreased after 40 GPa, that of sulfur-doped SnSe was stable for applying pressure at least 75.9 GPa. These differences would be considered by the carrier concentration of the sample crystal.

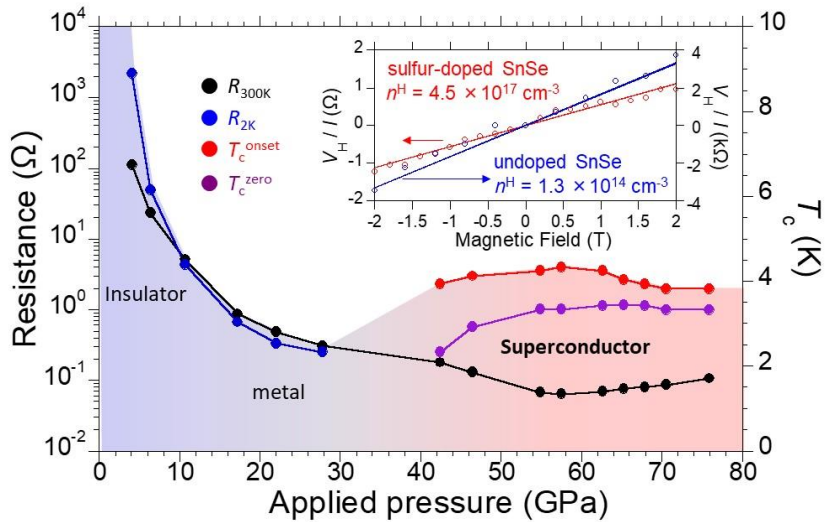

**Figure 6. Pressure phase diagram of SnSe$_{0.99}$S$_{0.01}$ up to 75.9 GPa. Inset shows magnetic field dependence of Hall voltage**

The inset of Fig. 6 shows Hall resistances as a function of magnetic field of the sulfur-doped SnSe and undoped SnSe single crystals at 300 K. From the slope of $V_H/I$ versus the magnetic field, the carrier concentrations of samples have been calculated using the formula, $(V_H/I) = (1/ned)H$, where $V_H$ is the Hall voltage, $I$ is the current, $n$ is the number of carriers, $e$ is the elementary charge, $H$ is the magnetic field and $d$ is the sample thickness. The figure shows that the sulfur-doped SnSe single crystal exhibits a positive slope for the $V_H/I$ versus $H$ curve, indicating a p-type characteristic with a carrier concentration of $4.5 \times 10^{17}$ cm$^{-3}$ at 300 K. This carrier concentration is dramatically higher than that of the undoped SnSe of $1.3 \times 10^{14}$ cm$^{-3}$. The higher carrier concentration of sulfur-doped SnSe may affect the pressure effect for the $T_c$.

## 4. Conclusion

The sulfur-doped SnSe single crystal was successfully synthesized in single crystals. The XPS analysis revealed the oxygen-sensitive surface state in the sample with valence fluctuation of $Sn^{2+}$ and $Sn^{4+}$. To investigate the high pressure effect for sulfur-doped SnSe, the pressure range of BDDE-DAC was upgraded to 104 GPa. The electrical resistivity measurement by using the DAC showed the transition from insulator to metal, and superconducting transition at higher pressure. $T_c^{onset}$ and $T_c^{zero}$ were observed around 4.3 K and 3.4 K, respectively. The Hall measurements showed that the hole carrier concentration of sulfur-doped SnSe is higher than that of the undoped SnSe. It was suggested the superconducting property under high pressure was affected by the carrier concentration.

**Acknowledgment**

This work was partly supported by JST CREST, Japan, JST-Mirai Program Grant Number JPMJMI17A2, Japan, and JSPS KAKENHI Grant Number JP17J05926. A part of the fabrication process was supported by NIMS Nanofabrication Platform in Nanotechnology Platform Project sponsored by the Ministry of Education, Culture, Sports, Science and Technology (MEXT), Japan. The part of the high pressure experiments were supported by the Visiting Researcher's Program of Geodynamics Research Center, Ehime University.